\documentclass [aip,pop, showpacs,reprint,pre, superscriptaddress]{revtex4-2}
\usepackage{graphicx}
\usepackage{textcomp}
\usepackage{psfrag}
\usepackage{float}
\usepackage{amsmath}
\usepackage{mathtools}
\usepackage{xcolor}
\usepackage{amssymb}
\usepackage{romannum}
\usepackage{url}
\usepackage{hyperref}
\usepackage{orcidlink}

\usepackage[us,12hr]{datetime}

\begin{document}
	
	\title{Effect of confinement anisotropy on particle transport and structural relaxation of finite dust clusters in complex plasma}
    
	\author{Hirakjyoti Sarma\,\orcidlink{0000-0003-0325-5650}}
    \email{hirakphy2019@gmail.com}
\affiliation{Institute for Plasma Research, Bhat, Gandhinagar-382428, India}
    
    \author{Sushree Monalisha Sahu\,\orcidlink{0009-0002-7263-702X}}

\affiliation{Institute for Plasma Research, Bhat, Gandhinagar-382428, India}
\affiliation{Homi Bhabha National Institute, Anushaktinagar, Mumbai 400094, India}

\author{P. Bandyopadhyay\,\orcidlink{0000-0002-1857-8711}}
\affiliation{Institute for Plasma Research, Bhat, Gandhinagar-382428, India}
\affiliation{Homi Bhabha National Institute, Anushaktinagar, Mumbai 400094, India}

\author{Ankit Dhaka\,\orcidlink{0000-0003-4425-9391}}
\affiliation{Atomic Semi Inc., Connecticut Street, San Francisco, CA 94107, USA}

\author{A. Sen\,\orcidlink{0000-0001-9878-4330}}
\affiliation{Institute for Plasma Research, Bhat, Gandhinagar-382428, India}
\affiliation{Homi Bhabha National Institute, Anushaktinagar, Mumbai 400094, India}

\date{\today ; \currenttime}
	\begin{abstract}
A finite dusty plasma cluster of charged microparticles confined in an anisotropic potential well is investigated experimentally and through Langevin dynamics simulations. As the confinement anisotropy is increased, the cluster undergoes a structural transition from near isotropic concentric shells to a linear chain configuration. The spatiotemporal modes of the cluster are analyzed by using Singular Value Decomposition. {At weaker confinement anisotropies, two modes are dominant. Mode 1, corresponding to a breathing-type oscillation of the cluster, and mode 2, representing an azimuthal rotational motion of the cluster, together carry around 99$\%$ of the signal energy.} With increasing anisotropy, the dominance of mode 2 decreases and that of mode 1 increases. This mode restructuring is accompanied by an increasingly non-Gaussian particle displacement statistics as evidenced by positive values of the Non-Gaussian Parameter maintained over an extended time duration. Simultaneously, the increasing dominance of mode 1 and suppression of mode 2 is accompanied by a significant slowing down of structural relaxation, with the cluster eventually exhibiting signatures of structural arrest. At weaker anisotropy, the experimentally measured dynamical quantities are very sensitive to the initial conditions which account for the discrepancy between the experiment and initial condition averaged simulation results for these observables. This study offers insight into the mechanisms underlying anomalous transport and structural relaxation in anisotropically confined many-body systems. 
	\end{abstract}
	
	
	\maketitle
	
	\section{Introduction}
    Anomalous transport in spatially confined systems has attracted significant attention from the researchers in the recent decades. A prominent example arises in the study of biological cells where the motion of tracer particles in the crowded intracellular environment often deviates from Brownian diffusion and exhibits anomalous characteristics \cite{hofling2013anomalous}. Of particular interest, is the underlying mechanism responsible for such anomalous transport in confined many body systems. To explain subdiffusive transport in these systems, normally the standard diffusion equation is replaced by fractional kinetic equations or continuous time random walk framework. The anomalous transport is characterized mainly by behaviours such as non-Gaussian probability distribution of the particle displacements, long-range temporal correlation and non-exponential decay of modes \cite{metzler2000random, metzler2004restaurant}. 
    
    Spatio-temporal patterns in physical or biological systems are often linked to the diffusion properties of the system. The emergent coherent structures — such as vortices, jets or lanes — can significantly alter particle trajectories and modify effective transport coefficients. Chaotic advection for example can enhance mixing and significantly affect diffusion of particles in the system \cite{jones1991enhancement}. Lagrangian coherent structures significantly affect material transport \cite{haller2015lagrangian}. In an active medium, the emergence of collective behaviour does have an impact on the transport of the constituent particles \cite{bechinger2016active}.

     To extract spatiotemporal structures from experimental signals measured at different space and time points, Singular Value Decomposition (SVD) provides an useful analytical technique. In this method, the spatiotemporal signal is assumed to be separable into spatial and temporal components, and the signal is decomposed into a set of orthogonal spatial and temporal modes. Several related variants of this approach are commonly used in the analysis of multivariate signals, including Proper Orthogonal Decomposition (POD) \cite{berkooz1993proper}, Biorthogonal Decomposition (BD) \cite{aubry1992spatio}, Principal Component Analysis (PCA) \cite{gewers2021principal}, and Karhunen–Loève Decomposition (KLD) \cite{webber1997karhunen}. 
     
    These techniques have previously been employed to extract coherent structures in turbulent flows \cite{berkooz1993proper} and for order reduction in mechanical systems \cite{kerschen2005method}. Their application in turbulence was mainly motivated by the observation that the apparant complexity in description of turbulent flows may be reduced to the analysis of a small number of dominant modes. They have proven useful in studying the symmetries and transformations of turbulent flows, as well as the dynamics and scaling of coherent structures \cite{sirovich1987turbulence}. Similar techniques have also been applied to the study of fluctuation phenomena in plasmas \cite{dudok1994biorthogonal} and to the identification of zonal flows in tokamaks \cite{van2014use}.
    
    Complex (dusty) plasma offer an ideal platform to study the transport properties of interacting many body systems in confined geometries. Owing to their much larger size and mass than the electrons and ions, time scale of the dust particle is much longer than the plasma species and hence the particle dynamics can be resolved very easily. In dusty plasmas, small number of particles can be trapped and manipulated by external perturbation to study different collective and single particle phenomena. Small number of particles under spatial confinement in dusty plasmas have been studied earlier by many researchers with respect to stucture, dynamics and phase transitions \cite{juan1998observation, juan1999structures, melzer2003mode, arp2005confinement, boning2008melting}. 

    A confined two-dimensional particle system can be made asymmetric by tuning the strength of the confining potential independently along the two spatial directions. Such mesoscopic systems have been investigated earlier both experimentally and numerically \cite{candido1998structure, apolinario2005structure, saint2002macroscopic, melzer2006zigzag}. It has been shown that confinement anisotropy induces particle rearrangements within the crystalline structure and significantly influences structural phase transitions. The ground state configurations of such a system were determined experimentally \cite{saint2002macroscopic}. In mesoscopic asymmetric superconducting disks, a transition from a multivortex phase to a giant-vortex state was observed theoretically \cite{schweigert1998vortex}. Furthermore, numerical studies have revealed the existence of multiple first- and second-order structural transitions as a function of the anisotropy parameter of the external confinement which are reflected in the normal mode frequencies \cite{apolinario2005structure}. Rancova et al. investigated the structural transitions of laterally compressed two-dimensional Coulomb clusters and demonstrated that these transitions are accompanied by distinct signatures in the specific heat capacity. They further showed that, under increasing lateral compression, the clusters evolve from a circular symmetric configuration to a linear chain through intermediate broken-symmetry states \cite{rancova2011structural}.
    
    Recently, it has been demonstrated that systematic variation of the channel width in the lower electrode of the Capacitively Coupled Dusty Plasma experimental (CCDPx) device \cite{Dhaka2025CapacitivelyPlasmas} modifies the plasma sheath, thereby generating an anisotropic confining potential whose strength can be controlled. This enables precise tuning of the confinement anisotropy without altering the background plasma parameters. Furthermore, structural transitions of the cluster are found to be driven solely by variations in the confinement anisotropy \cite{sahu2025confinement}. 
    While the study provides important insights into the equilibrium configurations of confined dust clusters, the associated transport phenomena emerging during such structural transitions remain largely unexplored. In particular, the spatiotemporal processes responsible for anomalous particle transport, collective rearrangements, and nonequilibrium energy redistribution in anisotropy-driven transitions are still not well understood.
    
    To address this gap, the Singular Value Decomposition (SVD) method provides a suitable framework for identifying spatiotemporal structures that lead to anomalous transport in confined many-body systems. Earlier studies have successfully employed this technique to analyze mode dynamics and nonequilibrium melting in finite dust clusters \cite{ivanov2009modes, ivanov2005melting}. SVD has been shown to serve as a complementary approach to conventional normal mode analysis, with the ability to uncover patterns in experimental data that may not be captured by analytical normal mode methods alone. 
    
    In the present work, we revisit the particle dynamics of a finite cluster of $N=12$ particles, as reported earlier by Sahu {\it et al.}\cite{sahu2025confinement} with systematically varying confinement anisotropy. We seek to elucidate the underlying mechanism responsible for the occurrence of anomalous transport in the finite cluster as the confinement anisotropy is tuned. To this end, we employ the method of Singular Value Decomposition (SVD) to extract the dominant spatial and temporal modes from the measured particle trajectories of the cluster. We further carry out Langevin Dynamics simulations and perform SVD analysis on the simulation trajectories which complement our experimental investigation.
    
     The remainder of the paper is organized as follows. Section \ref{Exp} contains the experimental set-up and procedures. In Section \ref{LD} the simulation strategy is explained. In section \ref{results} the results are discussed and section \ref{summary} contains the summary of the work.

    \section{Experimental set-up and procedures}\label{Exp}
The experiments were performed in the Capacitively Coupled Dusty Plasma experimental (CCDPx) device \cite{Dhaka2025CapacitivelyPlasmas}. Plasma is produced between two parallel circular electrodes driven by a 13.56~MHz RF source in push--pull mode at $\sim$3~W and 190~V$_{\mathrm{pp}}$ at a pressure of 1.5~Pa. An impedance matching network ensures efficient power transfer. RF-compensated Langmuir probe measurements \cite{Sudit1994RfDischarges} yield an electron temperature of 3--5~eV, plasma density of $(3$--$5)\times10^{15}$~m$^{-3}$, and plasma potential of 8--10~V for a range of discharge conditions.\\
A variable-width confinement channel is formed at the lower electrode using two D-shaped segments out of which, one is actuated by a stepper motor controlled via a microcontroller. The electrode separation can be tuned from 0 to $\sim$7.8~mm, enabling controlled variation of the in-plane confinement anisotropy without significantly altering bulk plasma parameters \cite{sahu2025confinement}. Monodispersive melamine--formaldehyde (MF) particles of diameter $7.43 \pm 0.05~\mu$m are introduced into the plasma. The particles acquire negative charge and levitate near the sheath edge where the vertical electric force balances gravity, forming a two-dimensional monolayer.
In the experiment, the confinement anisotropy was controlled by varying the channel width. The degree of anisotropy is quantified by the anisotropy parameter $\alpha$, defined as the ratio of the confinement strength along the $x$-direction to that along the $y$-direction, measured at the center of the channel. The parameter $\alpha$ was varied over the range $\sim 1$ to $0.004$. Larger values of $\alpha$ correspond to weaker anisotropic confinement, whereas smaller values indicate stronger confinement anisotropy.

The dust crystal is illuminated by a 630~nm, 350~mW red line laser in the horizontal plane and a 532~nm green line laser in the vertical plane. Particle dynamics are recorded simultaneously using top- and side-view high-speed CMOS cameras. The top-view camera provides a resolution of $1936 \times 1216$ pixels at up to 156~fps, while the side-view camera offers 12~MP resolution at up to 50~fps. The spatial resolutions of the top- and side-view imaging systems are 15~$\mu$m/pixel and 28~$\mu$m/pixel, respectively. Video data from both cameras are recorded for subsequent image analysis. The particle coordinates are obtained using Python based Trackpy package \cite{TrackpyPackage}.
 \begin{figure}
    \centering
    \includegraphics[width=0.9\linewidth]{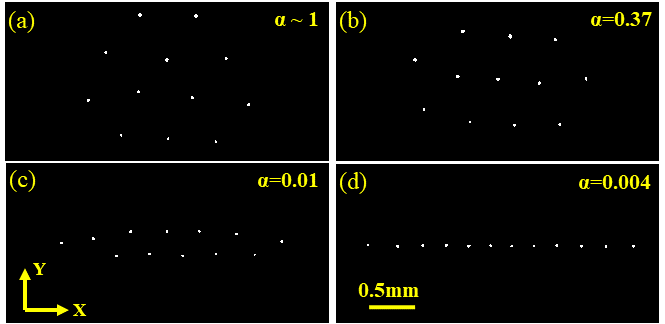}
    \caption{Experimentally obtained structures of twelve-particle cluster for (a) $\alpha\sim1$,  (b) $\alpha = 0.37$, (c) $\alpha = 0.01$ and (d) $\alpha = 0.004$.}
    \label{fig:structures}
\end{figure}
    \section{Langevin Dynamics Simulation}\label{LD}
	The dynamics of the two dimensional cluster consisting of $N=12$ charged dust particles in complex plasma is studied numerically via Langevin dynamics simulation using LAMMPS \cite{LAMMPS}. We assume the interaction among the particles to be governed by the screened Coulomb potential given by the following expression,
	\begin{equation}
		V_y(r_{ij}) = \frac{q_d}{4\pi \epsilon_0 r_{ij}}\exp(-r_{ij}/\lambda_d),
	\end{equation}
	where, $r_{ij}$ is the interparticle distance between the $i$th and $j$th particle, $q_d$ is the charge on the dust particles and $\lambda_d$ is the Debye length of the dust particles.
	 The charge and mass of the dust particles have been taken to be $q_d=10^4\;e$ and $m=3.24\times10^{-13}\;kg$ respectively. The experimentally measured confinement potential profiles suggest the presence of local Gaussian anharmonicity with a very weak global background of harmonic potential \cite{sahu2025confinement}. Therefore, in the simulation the particles are spatially confined by a two-dimensional potential given by the following expression,
	\begin{align}
		V_c(x,y) &= \frac{1}{2} K \Big[\Big(x-\frac{L}{2}\Big)^2 +\Big(y-\frac{L}{2}\Big)^2\Big] \nonumber \\ &+ A_x\exp\Big(-\frac{(x-\frac{L}{2})^2}{2\sigma_x^2}\Big ) + A_y\exp\Big(-\frac{(y-\frac{L}{2})^2}{2\sigma_y^2}\Big),
	\end{align}
	where, $L$ is the simulation box length. Near the trap center, the Gaussian potential can be approximated by a harmonic-oscillator potential. In this model, an anisotropy parmeter can be defined as, $\alpha=\frac{K-\frac{A_x}{\sigma_x^2}}{K-\frac{A_y}{\sigma_y^2}}$.
	We calculate the harmonic confining frequency as $\omega_0 = \sqrt{\frac{K q_d}{m}}$, 
	where, $K$ is the spring constant. 
	The Equation of Motion of the $i$th dust particle thus becomes,
	\begin{align}
		m\ddot{\mathbf{r}}_i &= -q_d\displaystyle \sum_{j\neq i}^{N}\mathbf{\nabla} V_y(r_{ij})\nonumber \\& - m\omega_0^2 \Big(x_i-\frac{L}{2}\Big) \mathbf{\hat{e_x}} -m\omega_0^2 \Big(y_i-\frac{L}{2}\Big) \mathbf{\hat{e_y}}\nonumber \\&+q_d\Big[\frac{A_x}{\sigma_x^2}\Big(x_i-\frac{L}{2}\Big)\exp\Big(-\frac{(x_i-\frac{L}{2})^2}{2\sigma_x^2}\Big )\mathbf{\hat{e_x}}\nonumber \\&+\frac{A_y}{\sigma_y^2}\Big(y_i-\frac{L}{2}\Big)\exp\Big(-\frac{(y_i-\frac{L}{2})^2}{2\sigma_y^2}\Big)\mathbf{\hat{e_y}}\Big]\nonumber \\&-m\nu_{dn}\dot{\mathbf{r}}_i+\mathbf{R_i}(t).
	\end{align}
	In the above equation, $\nu_{dn}$ represents the dust-neutral collision frequency and $\mathbf{R_i}$(t) is the random force on the $i$th particle which is assumed to have zero mean and a variance that is set according to the fluctuation-dissipation relation \cite{hansen2013theory}. 
	
	The system of dust particles in the simulation is first allowed to achieve a steady state at a set temperature $T_d$ by running the simulation for $5\times 10^5$ number of time steps. The trajectory data is collected for $5.6\times10^4$ number of time steps after the equilibration period is over. We keep the dust temperature fixed at $T_d=300\;K$ in the simulation. The dust neutral collision frequency is fixed as $\nu_{dn}=7.4\;s^{-1}$ which corresponds to the experimental value of neutral pressure $1.5$ Pa and the harmonic confinement strength is fixed as $\omega_0 = 0.04 \;rad\;s^{-1}.$ The timestep is fixed in the simulation as $\Delta t = 10^{-3}\;s.$

	\section{Results and Discussion}\label{results}
    
	\subsection{SVD of the Experimental Data}
\begin{figure}[htbp]
		\centering
		\includegraphics[width=0.9\linewidth]{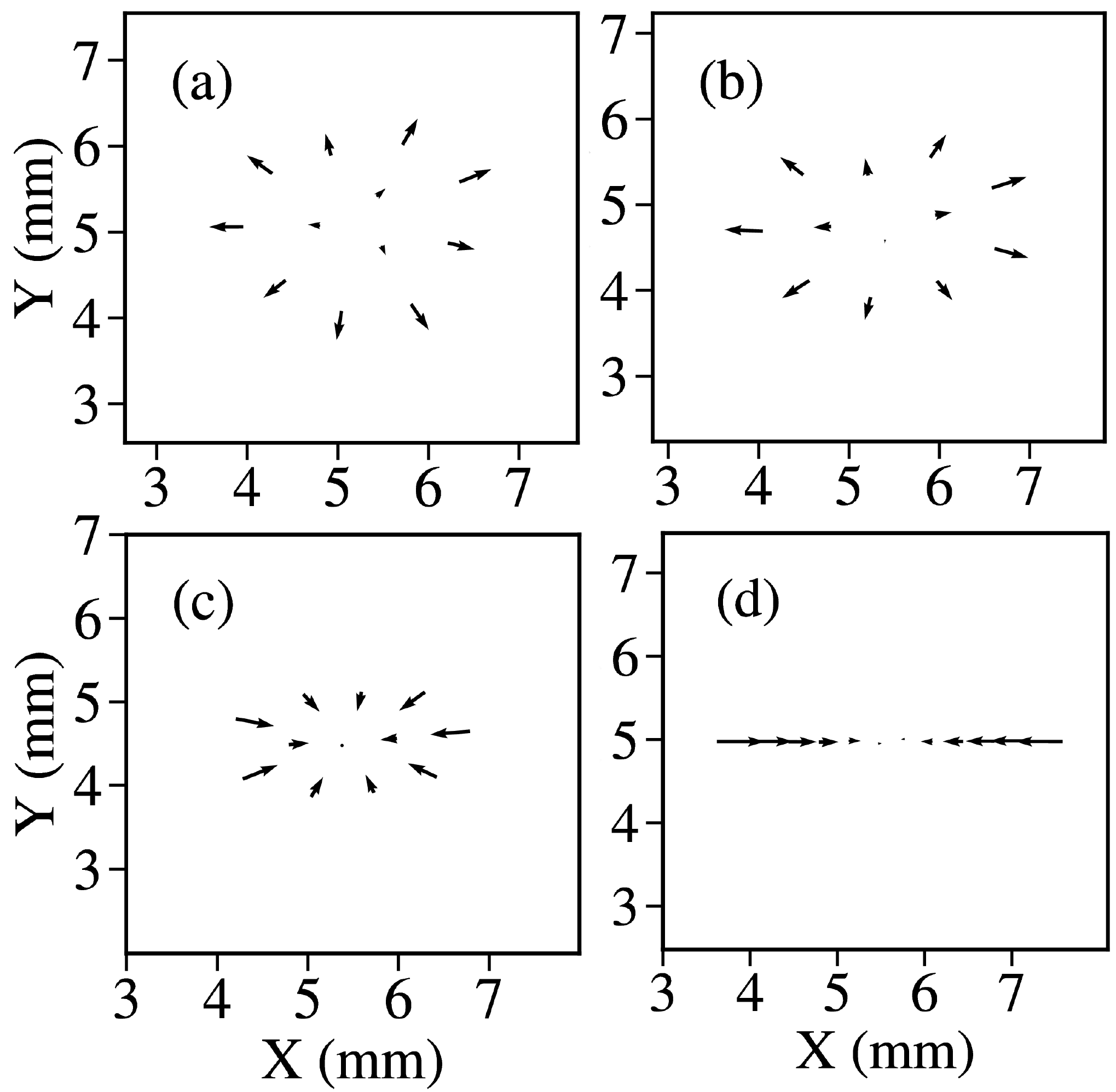}
    
		\caption{{\it Topo} of mode 1 obtained by performing SVD for four values of the confinement anisotropy, (a) $\alpha \sim 1$, (b) $\alpha = 0.37$, (c) $\alpha = 0.23$ and (d) $\alpha = 0.004$.}
	\label{fig1}
	\end{figure}

\begin{figure}[htbp]
        \centering
		\includegraphics[width=0.9\linewidth]{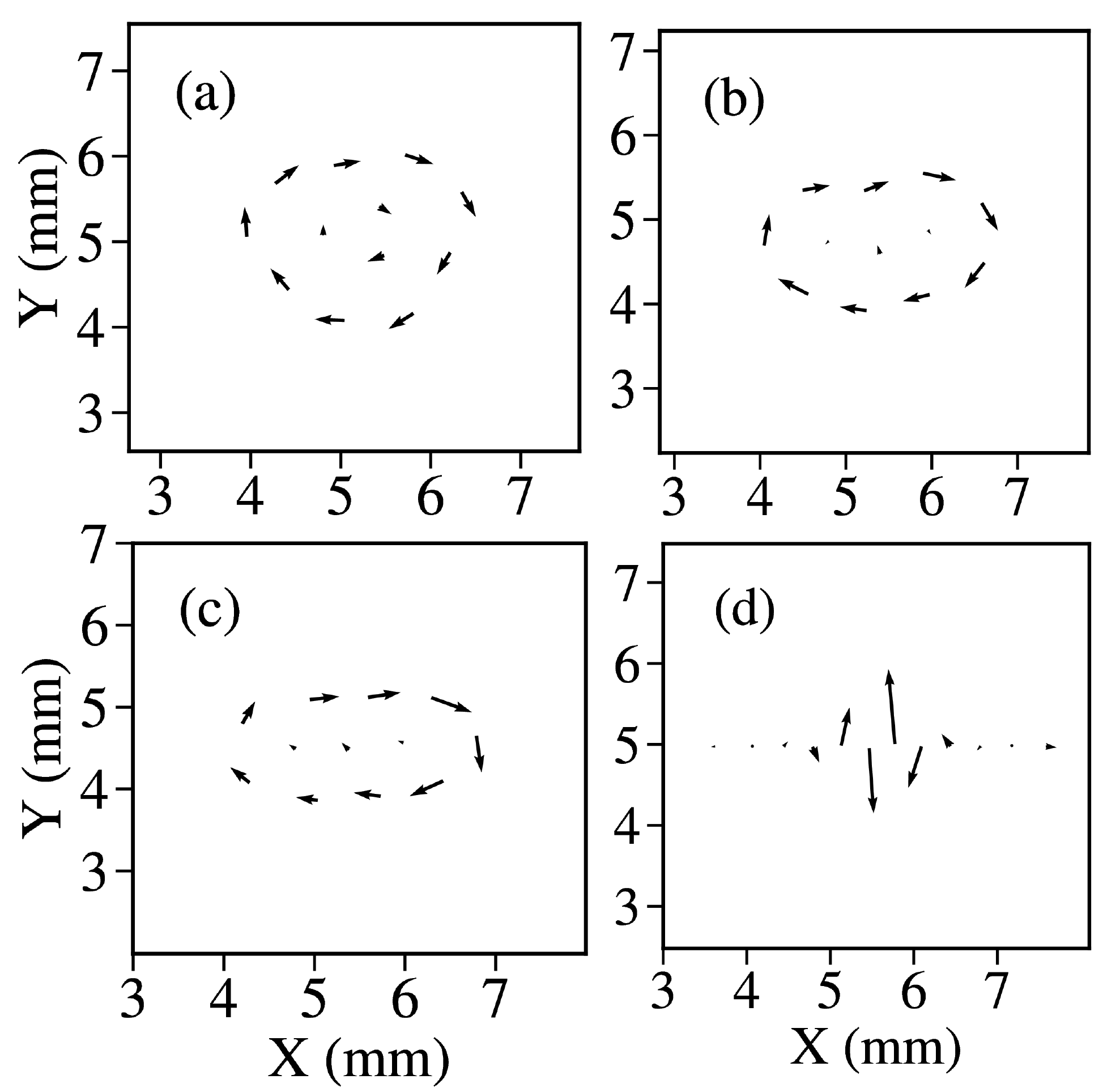}
    
		\caption{{\it Topo} of mode 2 obtained by performing SVD for four values of the confinement anisotropy, (a) $\alpha \sim 1$, (b) $\alpha = 0.37$, (c) $\alpha = 0.23$ and (d) $\alpha = 0.004$.}
		\label{fig2}
	\end{figure}

	The SVD decomposes the particle trajectories into orthongonal spatial and temporal modes. Although it doesn't necessarily correspond to the physical normal modes present in the system, for data which is sufficiently well sampled spatially and temporally it may correspond to the normal modes and is very sensitive to resonant linear normal modes \cite{van2014use}. For a brief discussion of SVD technique, the reader is referred to Appendix \ref{svd}. It is to be noted that, before performing SVD analysis we subtract a center-of-mass position from the data. A finite 2D charged particle cluster confined in a quadratic potential exhibits various normal modes such as rotational mode, breathing mode, vortex-antivortex mode etc \cite{schweigert1995spectral}. In the SVD framework, the spatial modes are called {\it topos.} The {\it topos} of mode 1 and mode 2 corresponding to some values of the anisotropy parameter are presented in Fig. \ref{fig1} and Fig. \ref{fig2}, respectively. At the weakest anisotropy, the topo of mode 1 reveals a breathing oscillation of the cluster similar to the breathing mode obtained in the normal mode analysis of a isotropic 2D cluster.   
    Similarly, at the highest value of the anisotropy parameter $\alpha = 1$ for which the cluster can reasonably be considered isotropic, the topology of mode 2 reveals a circulation similar to the rotational normal mode of a finite 2D cluster under isotropic harmonic confinement. But as the anisotropy of the confinement potential increases, the rotational mode is quenched and finally at the highest experimental anisotropy a new pattern emerges which is reminiscent of a standing wave in a 1D chain. At the lowest confinement anisotropy, the topology of mode 2 is longitudinal with respect to the circularly symmetric confinement geometry, whereas mode 1 exhibits a transverse character. As the anisotropy is increased and the confinement becomes elliptical, these modal characteristics are largely preserved. However, at the highest anisotropy, a clear exchange of modal character is observed. Mode 1 evolves into a pattern resembling the longitudinal acoustic mode of a linear chain, while mode 2 acquires features characteristic of a transverse optical mode. A similar behavior has been reported in the normal-mode analysis of anisotropically confined clusters interacting via a logarithmic potential, where the lowest-frequency mode undergoes an analogous evolution with increasing confinement anisotropy \cite{apolinario2005structure}. The observed exchange of modal character arises from the coupling between the two modes.  Similar behavior has been reported in quasi-one-dimensional chains subjected to a transverse magnetic field, where coupling between the acoustic and optical branches leads to an anti-crossing in the dispersion relation \cite{piacente2004generic}.
	
\begin{figure}[htbp]
		\centering
		\includegraphics[width=\linewidth]{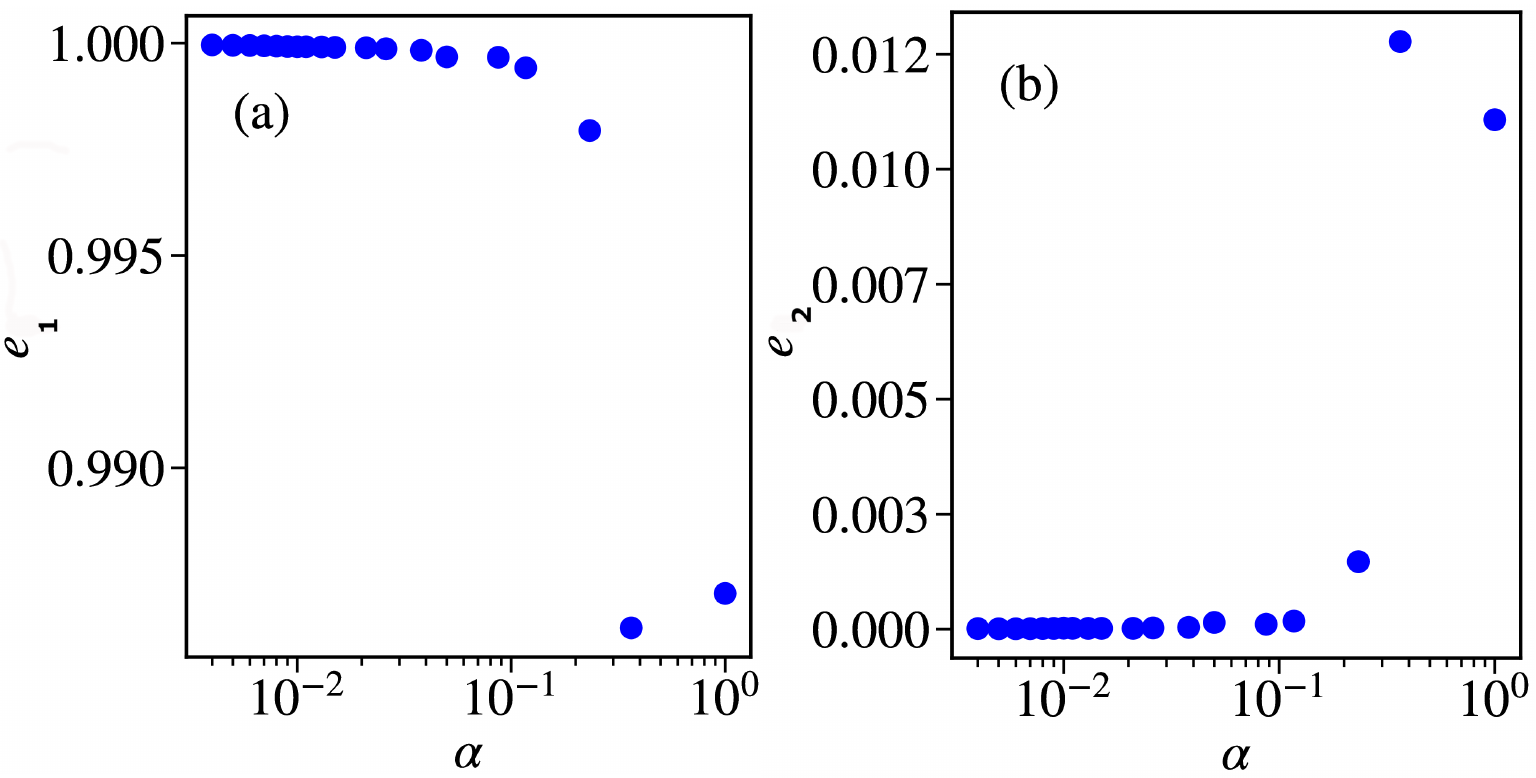}
    
		\caption{Relative weights of mode 1 (a) and mode 2 (b) as a function of the anisotropy parameter $\alpha$.}
		\label{rel_weights}
	\end{figure}

	 In order to assess the relative importance of mode 2 and mode 1 in the dynamics of the cluster, the relative weights of these two modes are plotted at different values of the anisotropy parameter in Fig. \ref{rel_weights}. It is to be noted that these two modes (mode 1 and 2) together carries around $99\%$ signal energy.
	{Although there is exchange of the character of the two modes, we see that the ordering of the relative weights remains the same, i.e, mode 1 posseses higher weight than mode 2 even at the highest anisotropy. An inspection of the amplitude of the topo of mode 1 and 2 for the linear chain reveals that at the higher anisotropy, mode 2 exhibits enhanced amplitude in the central region and reduced amplitude near the boundary, whereas mode 1 displays the opposite trend. The singular value corresponding to a mode is related to the variance contributed by that mode  to the total covariance of the signal measured at two different spatial locations while the topo determines the spatial distribution of that variance in the cluster\cite{van2014use}. Here, even if an exchange of modal character occurs between the two modes, the amplitude of the topos along the chain also changes in such a manner that the ordering of the relative weights remains preserved.}
    
    As can be seen from the Fig. \ref{rel_weights}, the relative weight shows a large change when the anisotropy parameter changes from $\alpha = 0.37$ to $\alpha = 0.23$. While the relative strength of the mode 2 sharply drops to zero with decreasing $\alpha$, the same for mode 1 quickly rises to unity. This indicates that there is an abrupt transfer of energy from the mode 2 to mode 1. The sharp drop of the relative weight of the mode 2 suggests a significant change in the dynamics of the particles in the cluster around this value of the anisotropy parameter.  
	
\begin{figure}[htbp]
		\centering
		\includegraphics[width=0.99\linewidth]{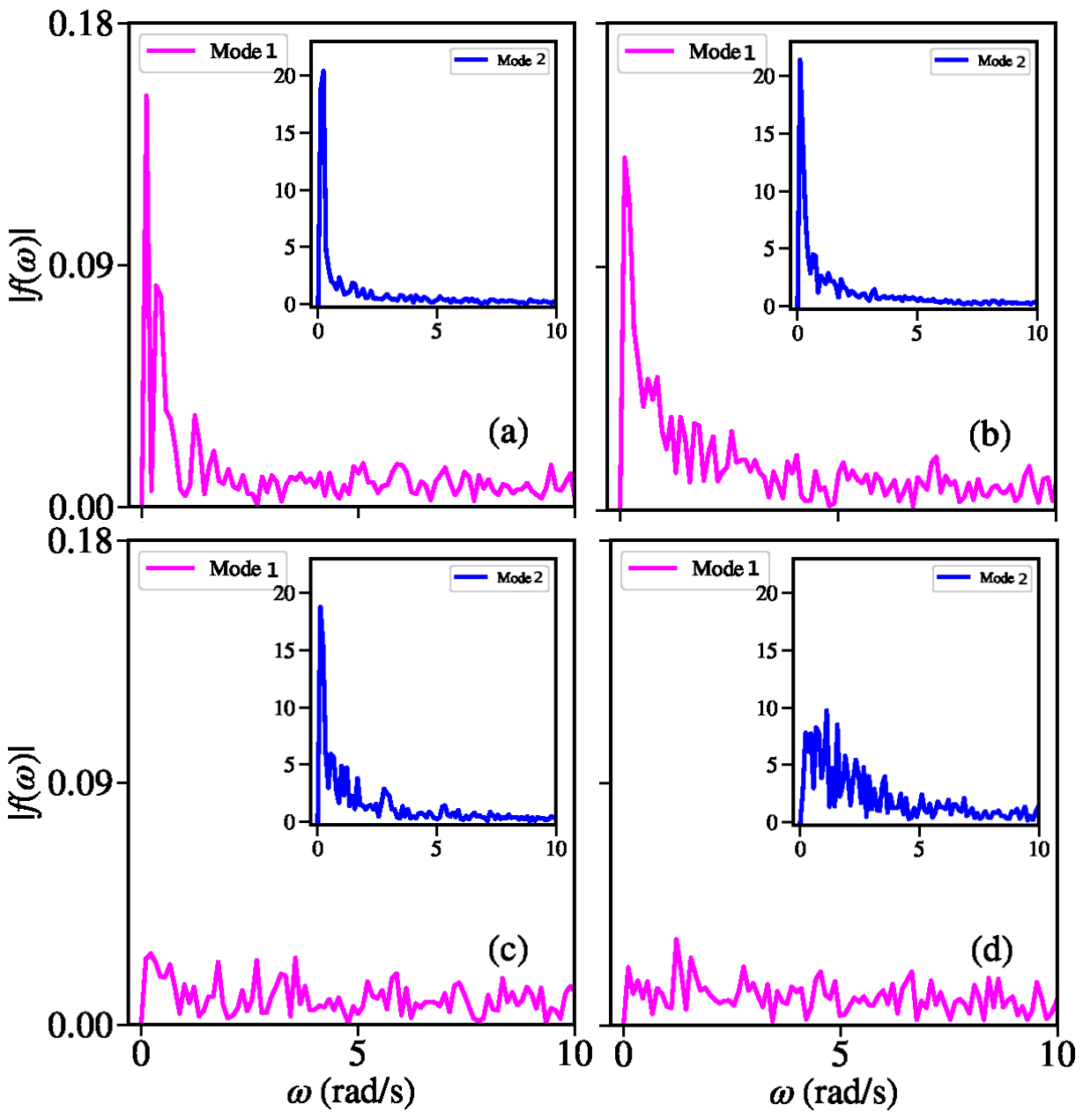}
		\caption{Fourier transform of the chrono of mode 1 and 2 at four different values of $\alpha$, (a) $\alpha \sim 1$, (b) $\alpha = 0.37$, (c) $\alpha = 0.23$ and (d) $\alpha = 0.1$.}
		\label{fft_chrono}
	\end{figure}

	The {\it chronos} which are contained in the matrix $V$ carries the time evolution of the {\it topos}. The time scale of the different modes can be inferred from the Fourier transform of the {\it chronos}. The Fourier transform of mode 1 and 2 at four different values of anisotropy parameter $\alpha$ is shown in Fig. \ref{fft_chrono}.
	The presence of a long lived coherent structure is indicated by a dominant peak in the Fourier spectrum of the chrono. It is seen from Fig. \ref{fft_chrono} that as the confinement anisotropy is increased, the coherent structure associated with mode 1 is suppressed at $\alpha=0.23$ [Fig. \ref{fft_chrono}(c)], as evidenced by the disappearance of the dominant spectral peak. Interestingly, despite the suppression of the spectral peak, the relative weight of mode 1 increases sharply at this value of $\alpha$ [Fig. \ref{rel_weights}(a)]. This suggests that the mode, although no longer associated with any coherent structure, still acquires greater importance in determining the overall dynamics of the system. Although, the coherent structure indicated by mode 1 is suppressed at this value of $\alpha$, mode 2 continues to exhibit signatures of coherent dynamics at $\alpha=0.23$, as indicated by the presence of a pronounced spectral peak. However,  the contribution of this mode to the overall dynamics remains weak, as the relative weight of mode 2 at $\alpha=0.23$ reduces significantly to  $\sim 0.002$ (Fig. \ref{rel_weights}(b)).
	
\begin{figure}[htbp]
        \centering
		\includegraphics[width=0.85\linewidth]{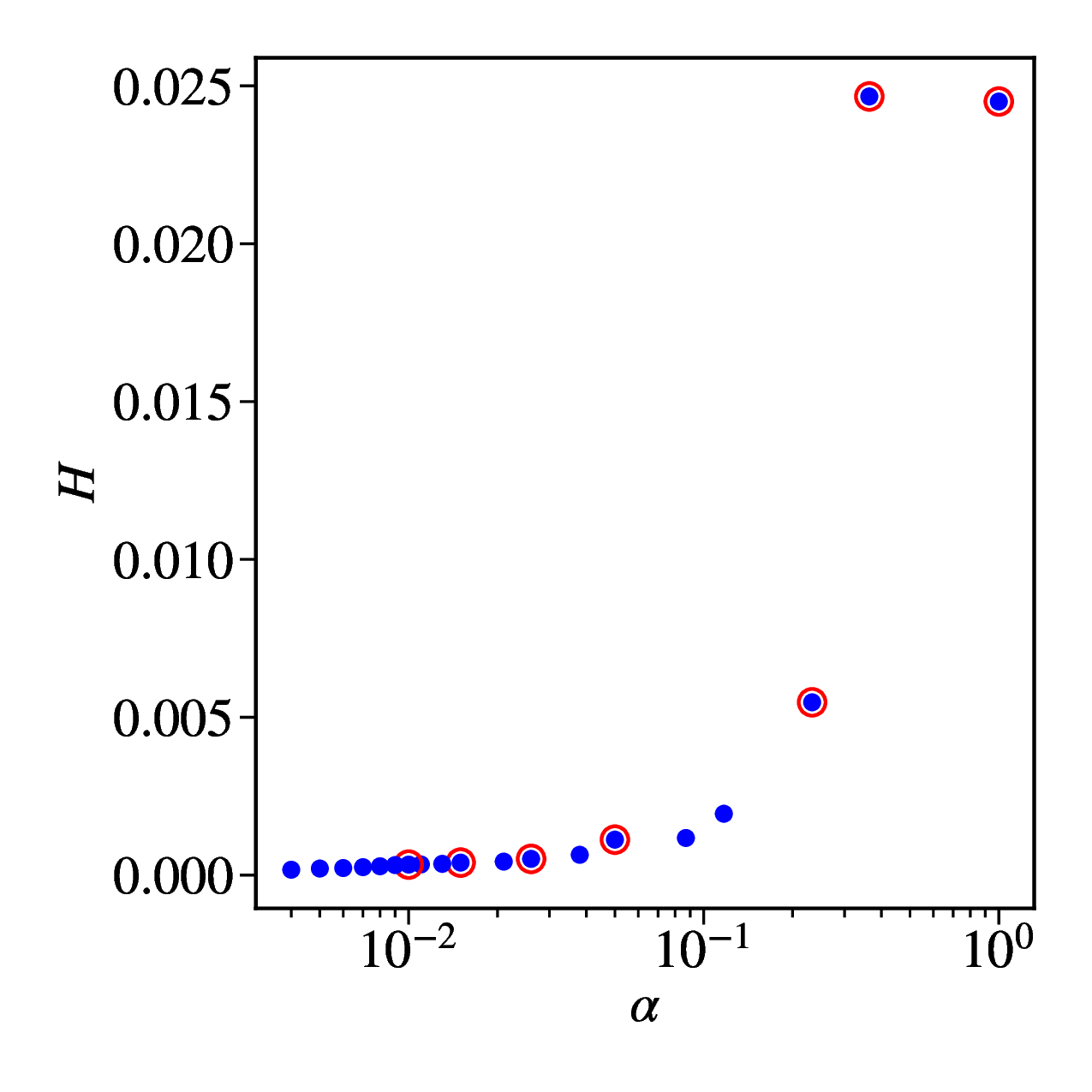}
		\caption{Normalized entropy of the anisotropically confined system of particles as a function of $\alpha$. The Mean Squared Displacement are plotted for the $\alpha$ values corresponding to the the red circled data points.}
		\label{norm_en}
	\end{figure}

	Another quantity which captures the space-time complexity is the normalized entropy of the system defined as \cite{dudok1994biorthogonal}, 
	\begin{equation}
		H = \frac{-\displaystyle \sum_{i=1}^{P} e_i \log{e_i}}{\log{P} },
	\end{equation}
	where, $P$ is the total number of SVD modes. The normalized entropy of the system is plotted in Fig. \ref{norm_en} as a function of the anisotropy parameter.
    The normalized entropy reaches unity if all the modes equally contribute to the dynamics (i.e, all singular values are equal), and becomes zero if the dynamics is completely described by only a single mode (i.e, only one singular value is  non-zero). From Fig. \ref{norm_en} it is seen that, for higher anisotropy ($\alpha < 0.1$) $H$ lies in the range $10^{-4}-10^{-3}$. As can be seen from the relative weight in Fig. \ref{rel_weights}(a), at the highest anisotropy the dynamics is almost entirely dominated by mode 1. The normalized entropy increases gradually upto $\sim 0.006$ over $0.1 < \alpha \leq 0.23$, before rising sharply to $\sim 0.025$ at $\alpha = 0.37$. At the lowest anisotropy ($\alpha \sim 1$), both mode 1 and mode 2 contribute significantly, although dominance of mode 1 remains larger. Nevertheless, 
    $H$ remains small compared to unity throughout, indicating the dynamics stays effectively low-dimensional across the full range of $\alpha$
     studied.
	\subsection{Transport of particles in the cluster}
\begin{figure}[htbp]
		\centering
        \includegraphics[width=0.95
        \linewidth]{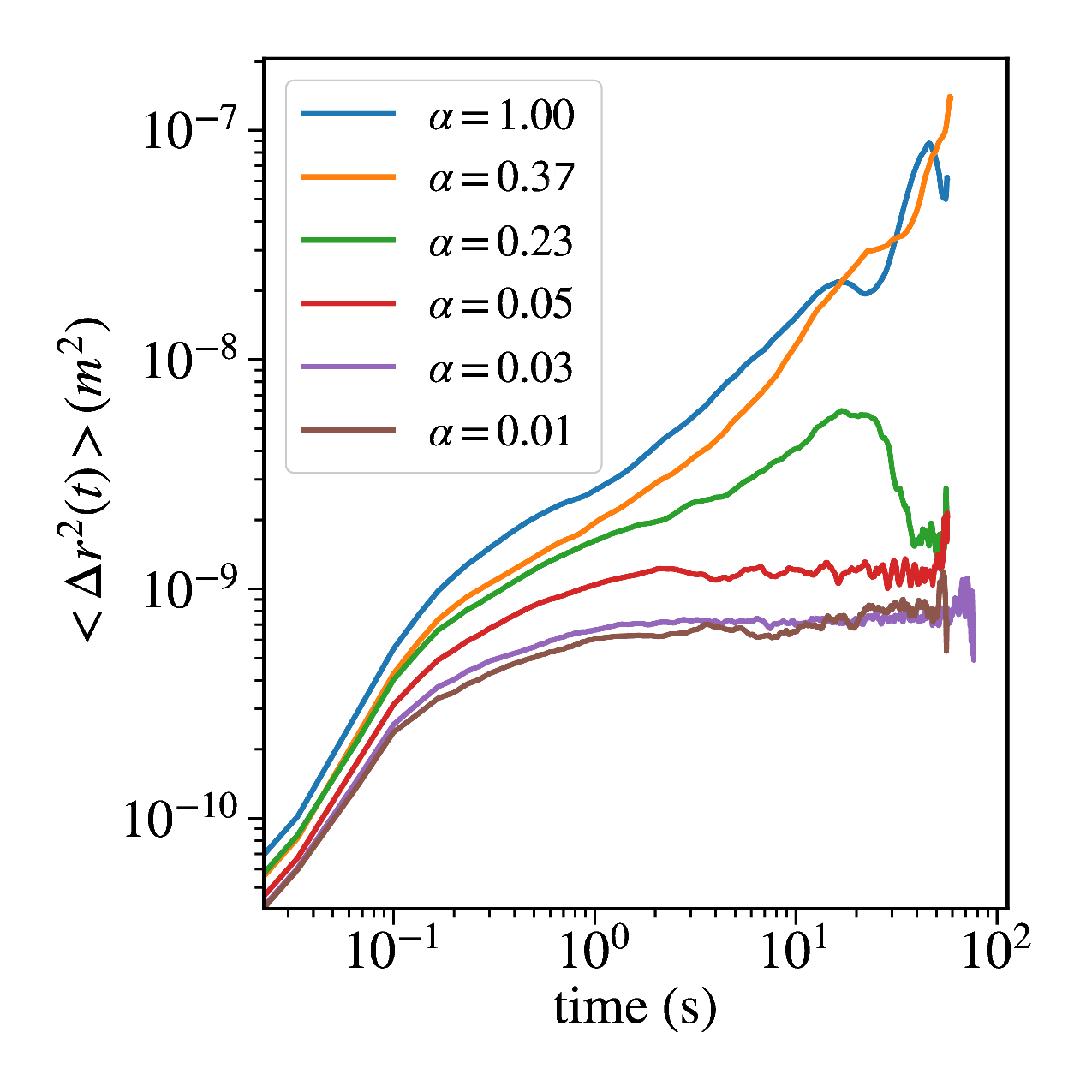}
		
		\caption{(Color online) Mean Squared Displacement of the system of particles at different values of $\alpha$ as a function of time.}
		\label{msd_expt}
	\end{figure}

	The different modes present in a finite cluster are known to have effects on the transport of particles in the cluster \cite{schella2013transport}. Therefore, the Mean Squared Displacement (MSD) is plotted as a function of time in Fig. \ref{msd_expt} for six different values of $\alpha$. The MSD is calculated according to the formula,
    \begin{equation}
        \langle \Delta r^2(t) \rangle = \frac{1}{N\tau_{max}}\sum_{\tau=1}^{\tau_{max}}\sum_{j=1}^{N}\left[\mathbf{r}_j(t+\tau)-\mathbf{r}_j(\tau)\right]^2,
    \end{equation}
    where, $\tau_{max}$ and $t$ denote the maximum number of time origins and delay time, respectively \cite{allen2017computer}. Thus, this is a both Time- and Ensemble-Averaged Mean Squared Displacement (TEAMSD) \cite{bewerunge2016time}.
	For a larger value of $\alpha$ (for which the cluster is resonably isotropic) the MSD can be seen to increase ballistically at short times and then showing oscillations at later times. At very short times particles are in ballistic regime, for which $\langle \Delta r^2(t) \rangle \approx \langle v^2 \rangle t^2.$ Since $\langle v^2 \rangle$ is proportional to kinetic temperature through the equipartition theorem, the magnitude of short time MSD scales with temperature. As shown by Sahu et al. \cite{sahu2025confinement}, the temperature of the system of particles changes with change in the confining anisotropy. This variation is also reflected in the MSDs shown in Fig. \ref{msd_expt}, where the MSDs can be seen to start from different values at different $\alpha$. The quenching of mode 2  and increasing dominance of mode 1 are accompanied by a  decrease in the mobility of the particles, as evidenced from the drop of the MSD at $\alpha=0.23$.
	
\begin{figure}[htbp]
		\centering
        \includegraphics[width=\linewidth]{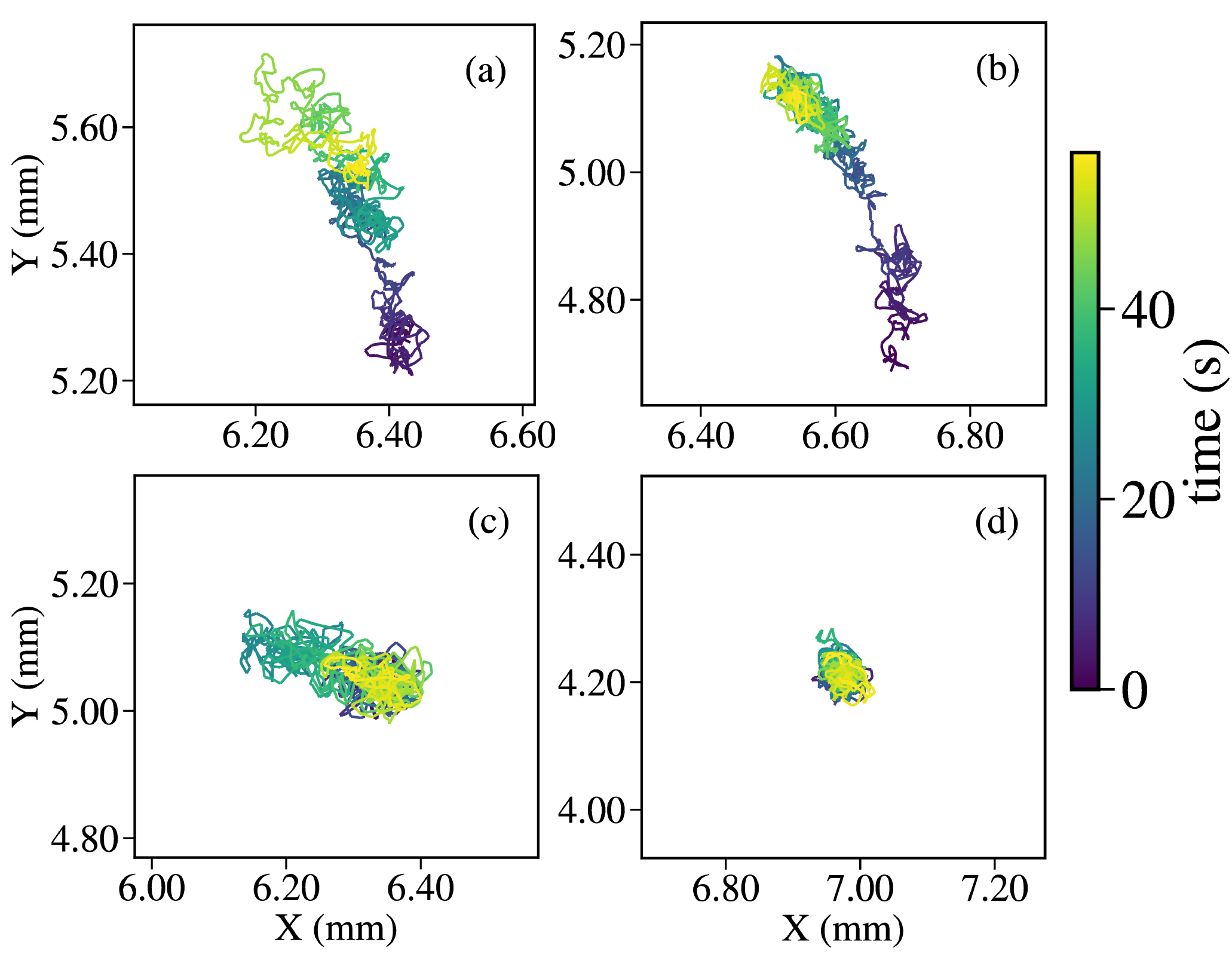}
		\caption{(Color online) Trajectory of a single particle from the outer portion of the cluster at  four different values of $\alpha$, (a) $\alpha = 1$, (b) $\alpha = 0.37$, (c) $\alpha = 0.23$ and (d) $\alpha = 0.1$.}
    \label{trajectory_single}
	\end{figure}

\begin{figure}[htbp]
		\includegraphics[width=0.95\linewidth]{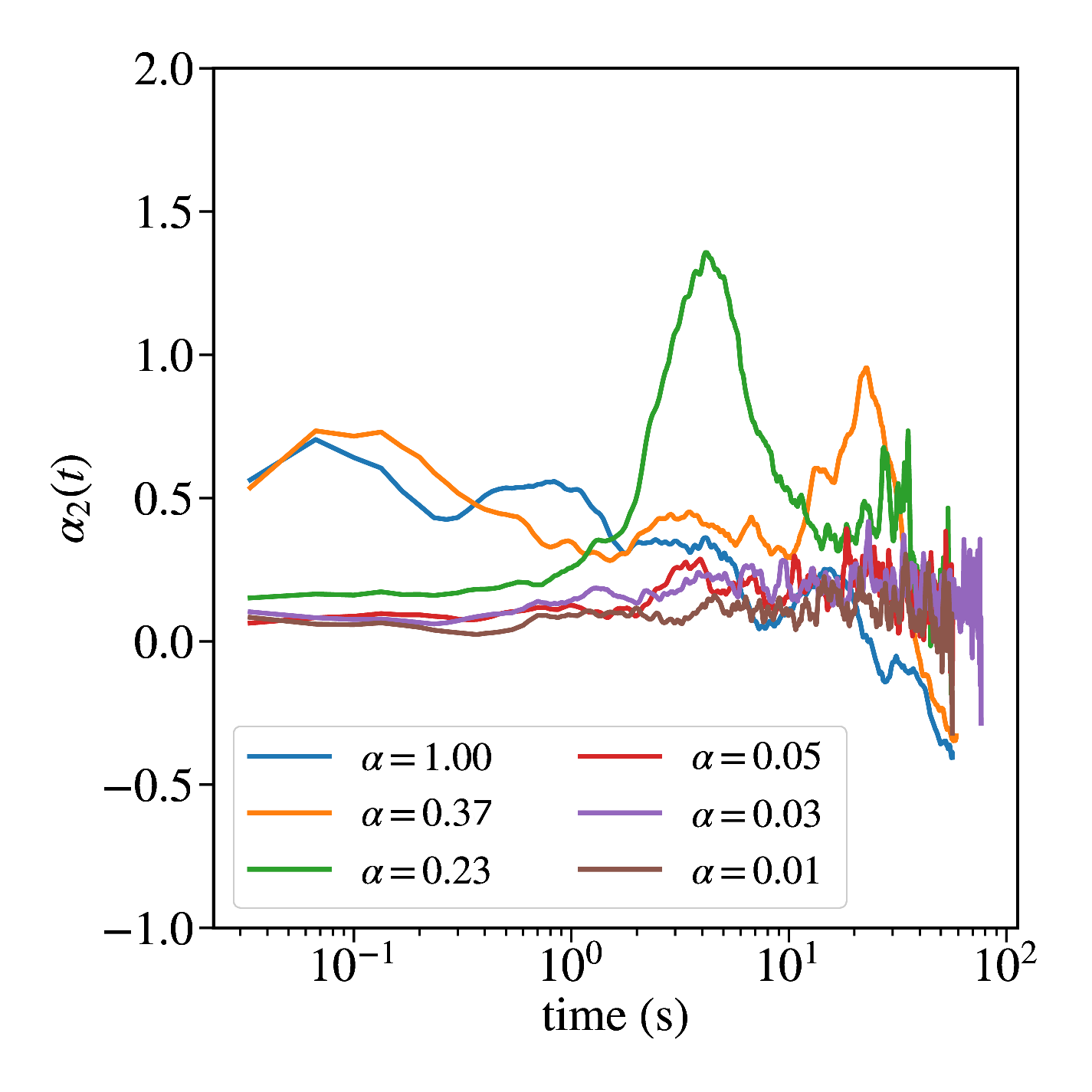}
		\caption{(Color online) Non-Gaussian parameter of the system of particles at different values of the anisotropy parameter as a function of time.}
		\label{ngp_expt}
	\end{figure}

	The trajectory of a typical outer layer particle for four values of the anisotropy parameter is shown in Fig. \ref{trajectory_single}.
	It is seen that for $\alpha = 1$ and $\alpha = 0.37$ the particle exhibits large jumps, whereas for smaller values of $\alpha$ the jump motion is suppressed.  This observation motivates us to find the non-Gaussian parameter which is shown in Fig. \ref{ngp_expt}. It is defined as,
    \begin{equation}
        \alpha_2(t) = \frac{\big<\Delta r^4(t)\big>}{2\langle \Delta r^2(t) \rangle}-1.
    \end{equation}
	The non-Gaussian Parameter (NGP) for diffusive transport tends to zero at longer times. However, when the transport is dominated by large jumps, the NGP exhibits a pronounced positive peak at the corresponding timescale before eventually decaying toward zero at longer times. For $\alpha = 1.0$, the NGP starts from a large positive value at short times and shows an overall decreasing trend, eventually attaining negative values at longer times. A similar overall behavior is observed for $\alpha = 0.37$, although in this case the NGP exhibits a temporary increase at intermediate times before continuing its decay toward negative values. For still lower values of $\alpha$, the NGP remains positive over both short and intermediate times; however, unlike the higher values of $\alpha$, it does not display a pronounced positive peak at intermediate times. 

   The decrease in MSD as the confinement anisotropy increases hints at the slowing down of the structural relaxation of the system of particles. Therefore, we obtain the autocorrelation of microscopic density at varying anisotropy of the confining potential. The density autocorrelation in wavenumber space is defined as, $F(\mathbf{k},t)=\Big<\rho(\mathbf{k},t)\rho(-\mathbf{k},0)\Big>$, where, $\displaystyle \rho(\mathbf{k},t)=\frac{1}{\sqrt{N}}\sum_{j=1}^{N}\exp(-i\mathbf{k}\cdot\mathbf{r}_j(t))$ is the spatial Fourier transform of the microscopic particle density \cite{boon1991molecular}. Here, the angle bracket $\Big<...\Big>$ denotes an average over a large number of initial times (time origins). In the field of neutron scattering, $F(\mathbf{k},t)$ is commonly known as the Intermediate Scattering Function (ISF). The component of the ISF that describes the correlation of a particle with itself over time is called the Self-Intermediate Scattering Function (SISF), and is given by, $F_S(\mathbf{k},t)=\frac{1}{N}\Big<\displaystyle \sum_{j=1}^{N}\exp(-i\mathbf{k}\cdot(\mathbf{r}_j(t)-\mathbf{r}_j(0))\Big>$.
   
    $F_S(\mathbf{k},t)$ gives an idea about the relaxation dynamics of a system of particles.
    For a simple liquid, in the long-time limit, $F_S(\mathbf{k},t)$ decays to zero exponentially. However, for a viscous melt, it exhibits a fast relaxation process ($\beta$-relaxation) followed by a slower relaxation ($\alpha$-relaxation), which is characterized by a stretched exponential decay. In the glassy state, structural relaxation is arrested, and at long times, SISF approaches a finite value called the non-ergodicity parameter \cite{ruta2017relaxation}.

\begin{figure}[htbp]
		\includegraphics[width=0.9\linewidth]{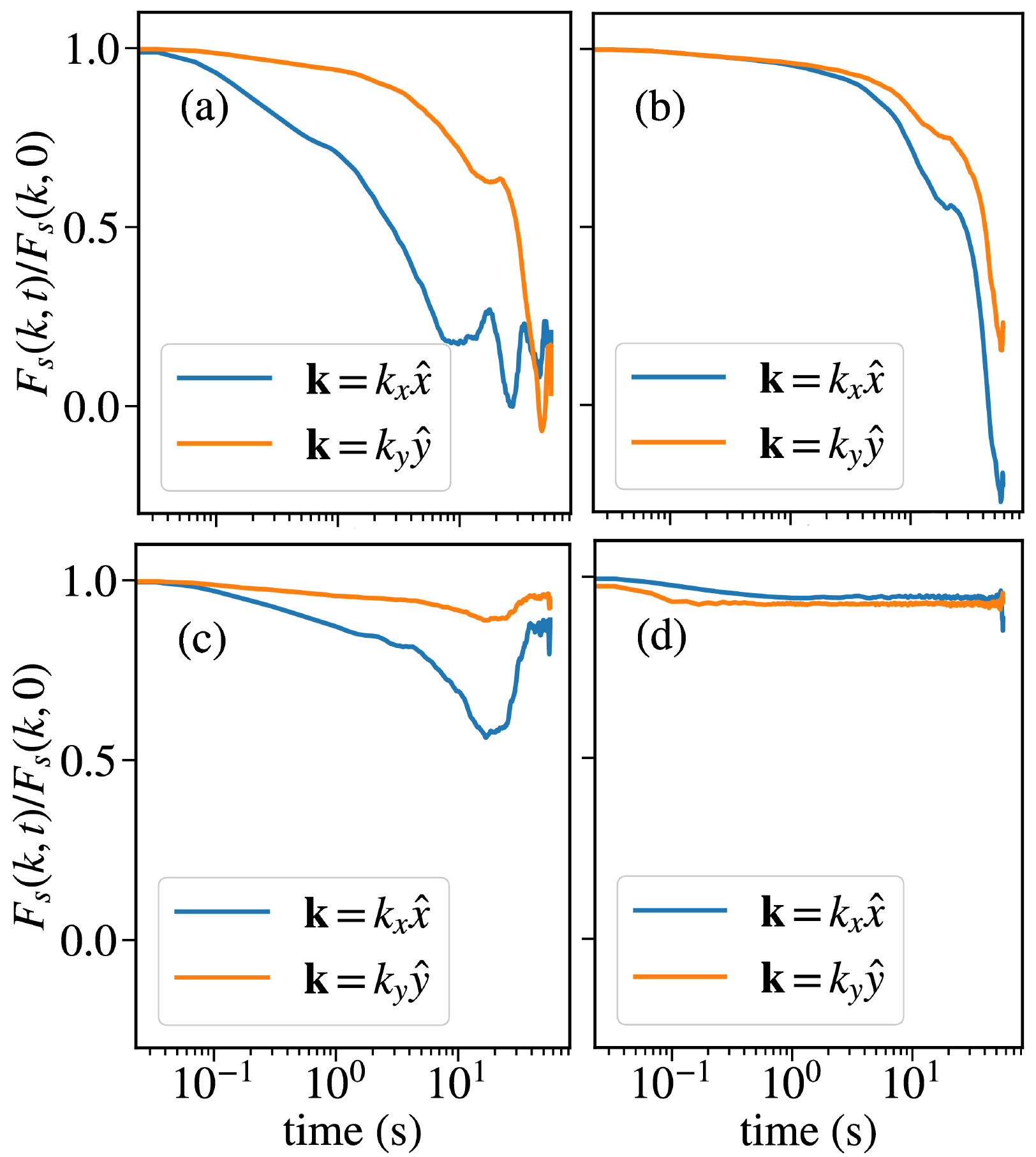}
		\caption{(Color online) Self-Intermediate Scattering Function at four values of the anisotropy parameter $\alpha$, (a) $\alpha \sim 1$, (b) $\alpha=0.37$, (c) $\alpha=0.23$ and (d) $\alpha=0.01$, exhibiting slowing down of the relaxation dynamics with increasing anisotropy.}
		\label{sisf_expt}
	\end{figure}

    Since our system is under an anisotropic confinement, we choose to plot $F_S(\mathbf{k},t)$ along the $x-$ and $y-$ directions separately ( Fig. \ref{sisf_expt}). The wavenumbers along $x-$ ($k_x$) and $y-$directions ($k_y$) are chosen to correspond to the first peak of the static structure factor in the respective directions, as this gives the nearest-neighbor length scale where structural relaxation is most pronounced. It is seen from Fig. \ref{sisf_expt} that there is a two-stage relaxation process present in the cluster when the confinement anisotropy is relatively weaker (Fig. \ref{sisf_expt}(a) and \ref{sisf_expt}(b)). When the anisotropy is increased, for $\alpha=0.23$ (Fig. \ref{sisf_expt}(c)), the SISF tend to decay initially, but then it rises again. In glass-forming systems, in the $\beta$-relaxation period the particles move within the cages formed by their neighbors (caging regime) whereas during the much slower $\alpha$-relaxation period particles are able to decage and exhibit diffusive motion (diffusive regime) \cite{feng2010identifying}. In Fig. \ref{sisf_expt}(c), the SISF exhibits an initial decay followed by a subsequent increase, without reaching zero. This suggests that the system undergoes only partial structural relaxation and retains significant memory of its initial configuration. This behaviour of SISF at $\alpha=0.23$ is accompanied by an abrupt transfer of signal energy from SVD mode 2 to SVD mode 1 (cf. Fig. \ref{rel_weights}) at that value of $\alpha$. On further decreasing the anisotropy parameter to $\alpha=0.01$ (Fig. \ref{sisf_expt}(d)) the cluster appears to undergo structural arrest in the long time limit. While the cluster exhibits slow structural relaxation with increasing anisotropy that is phenomenologically similar to that of a glass-forming system, this behavior is not identical to true glass formation in bulk disordered systems, and we do not claim the underlying physics to be equivalent. 
	\subsection{Comparison with Langevin dynamics simulations}
\begin{figure}[htbp]
		\centering\includegraphics[width=\linewidth]{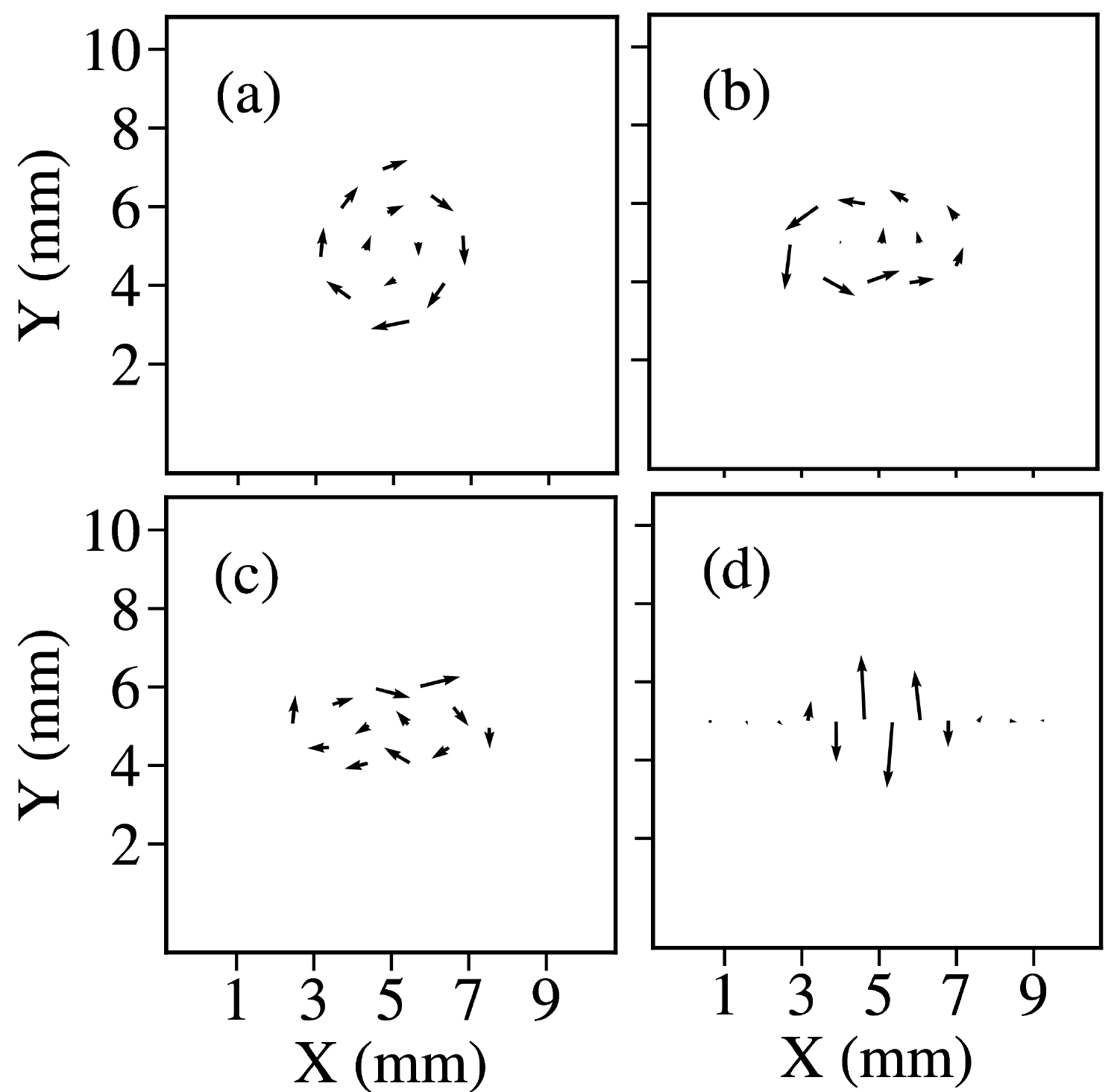}
		\caption{{\it Topo} of mode 2 obtained by performing SVD on the simulation data for four values of the confinement anisotropy, (a) $\alpha = 1$, (b) $\alpha = 0.36$, (c) $\alpha = 0.23$ and (d) $\alpha = 0.01$.}
		\label{sim_rot}
	\end{figure}

\begin{figure}[htbp]
		\centering\includegraphics[width=0.95\linewidth]{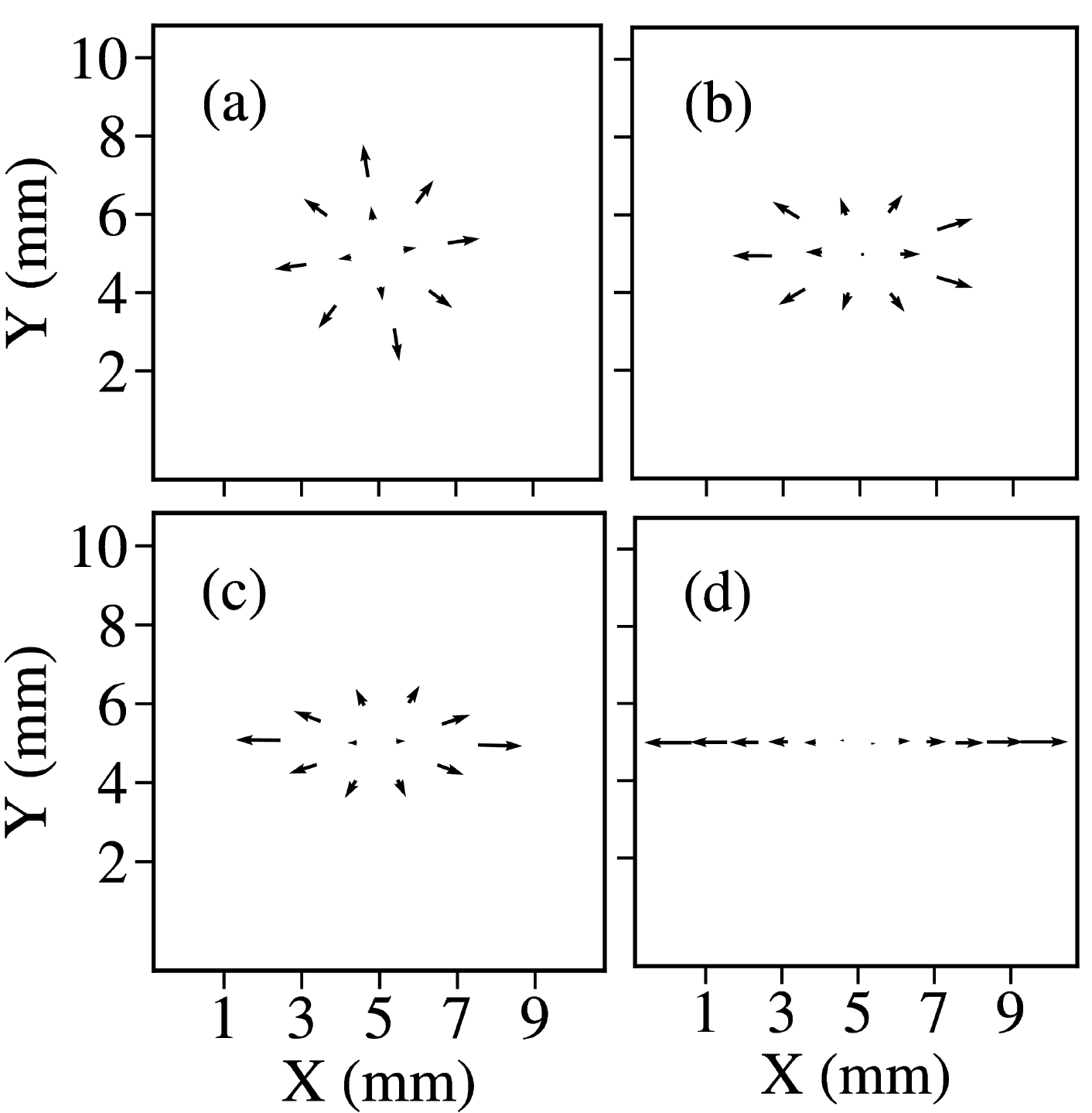}
		\caption{{\it Topo} of mode 1 obtained by performing SVD on the simulation data for four values of the confinement anisotropy, (a) $\alpha = 1$, (b) $\alpha = 0.36$, (c) $\alpha = 0.23$ and (d) $\alpha = 0.01$.}
		\label{sim_breath}
	\end{figure}

	To understand the experimental results, we compare the results with those from Langevin Dynamics simulation. We first present the spatial patterns of mode 2 and 1 obtained from the simulation for four different values of $\alpha$ as shown in Fig. \ref{sim_rot} and Fig. \ref{sim_breath}, respectively. The anisotropy parameter $\alpha$ is varied in the simulation by varying the parameter $\sigma_y$ while keeping $\sigma_x$ fixed.
	Similar to the experimentally obtained results (cf. Fig. \ref{fig1}) we see a circulation for mode 2 at larger values of the anisotropy parameter $\alpha$ and a breathing oscillation corresponding to mode 1. At sufficiently higher confinement anisotropy, where the cluster transforms into a linear chain, mode 1 exhibits transverse oscillations, whereas mode 2 becomes longitudinal with respect to confining geometry, consistent to the experimental observations.
    
\begin{figure}[htbp]
		\includegraphics[width=0.9\linewidth]{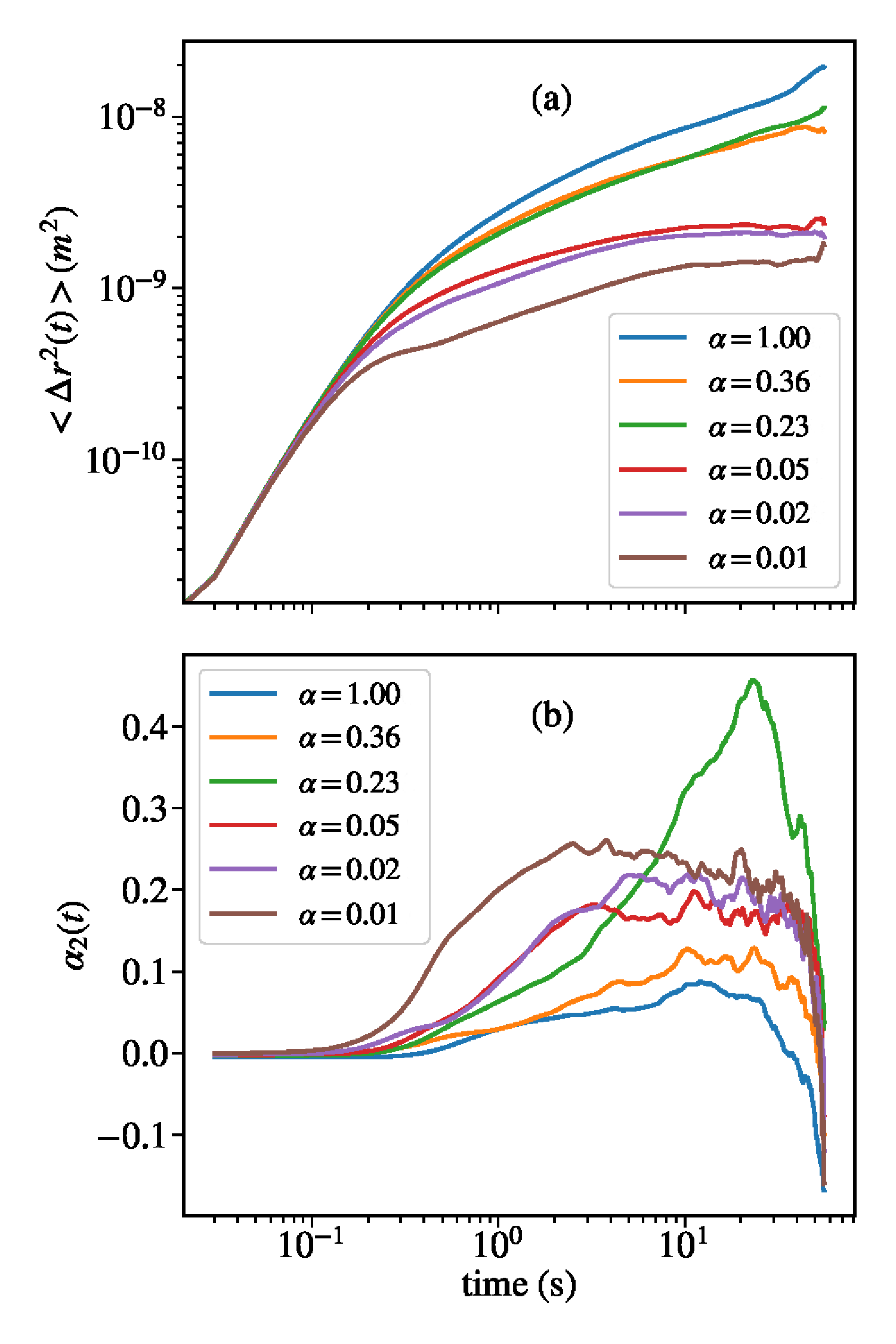}
		\caption{(Color online) Mean Squared Displacement and Non-Gaussian Parameter of the system of particles at different values of the anisotropy parameter as a function of delay time obtained from Langevin dynamics simulation.}
		\label{msd_ngp_sim}
	\end{figure}

	To study particle transport as a function of $\alpha$, we also obtain the Mean Squared Displacement (MSD) and the Non-Gaussian Parameter (NGP) from the simulation. We have obtained the MSD and NGP by performing average over a number of random independent initial conditions to ensure that they reflect the physical modes rather than fluctuations arising from specific initial conditions. The MSD and NGP are shown in Fig. \ref{msd_ngp_sim}.
	
\begin{figure}
        \centering
        \includegraphics[width=1\linewidth]{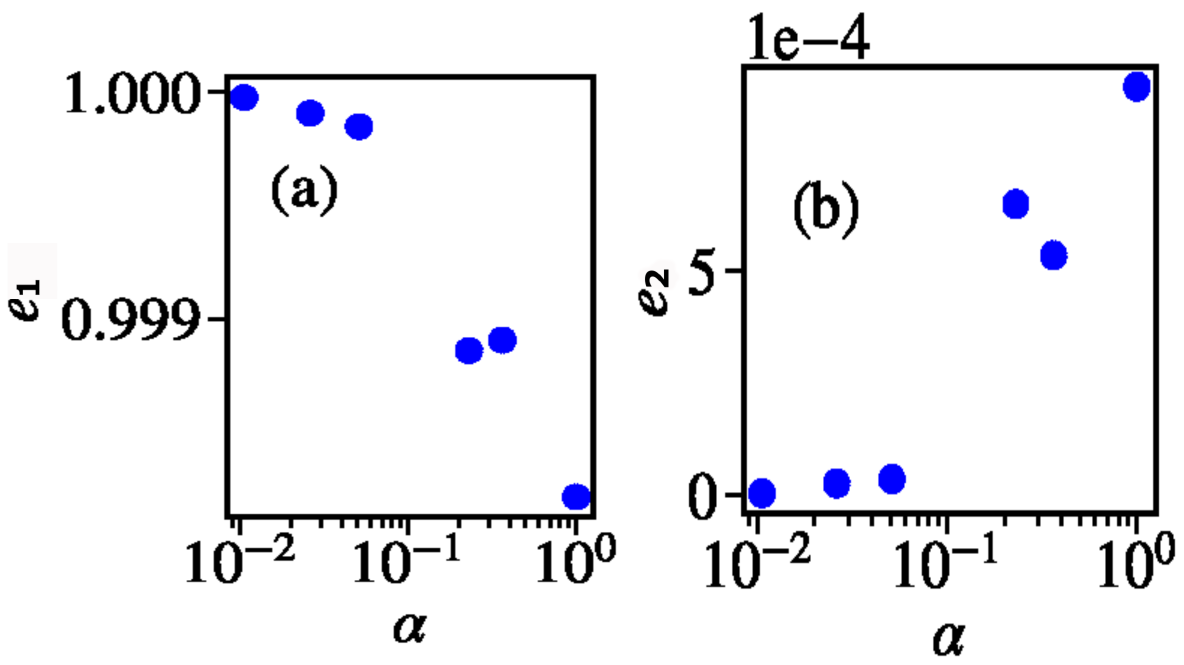}
        \caption{The relative weights of (a) mode 1  and (b) mode 2 as a function of $\alpha$ obtained from the Langevin Dynamics simulation.}
        \label{sing_sim}
    \end{figure}

	It can be seen that, as $\alpha$ decreases from unity, beyond a certain value of $\alpha$ ($\alpha=0.23$), there is a significant drop in the slope of MSD and a peak appears in the NGP. This indicates that at this value of $\alpha$, the transport of particles in the cluster undergoes a significant change. The relative weights averaged over initial conditions corresponding to these values of $\alpha$ for modes 1 and 2 are shown in Fig. \ref{sing_sim}.
    It is seen that the relative weight of mode 1 and 2 exhibits discrete change at $\alpha=0.23.$ The change that is observed experimentally while going from $\alpha=0.37$ to $\alpha=0.23$ is similar to the change observed in the simulation while going from $\alpha=0.23$ to $\alpha=0.05$ in the transport properties of the dust cluster. Also one can see the rise of NGP from nearly zero to large positive values with decrease in $\alpha$ from $1$ to $0.01$, i.e., with rise in confinement anisotropy as shown in Fig. \ref{msd_ngp_sim}(b) with a peak at an intermediate time scale for $\alpha=0.23$. 

\begin{figure}[htbp]
		\includegraphics[width=\linewidth]{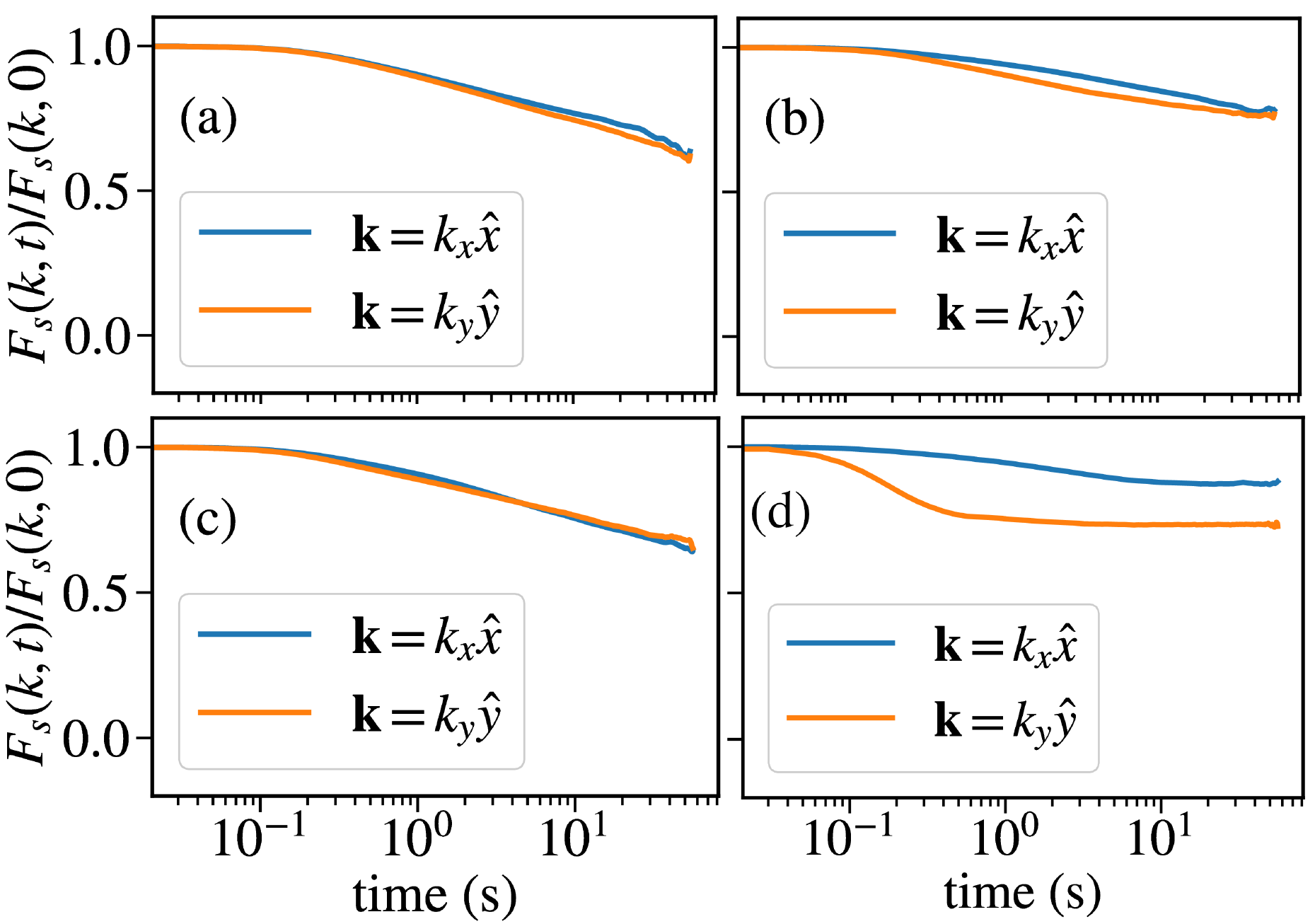}
		\caption{(Color online) Self-Intermediate Scattering Function at four values of the anisotropy parameter $\alpha$, (a) $\alpha= 1$, (b) $\alpha=0.36$, (c) $\alpha=0.23$ and (d) $\alpha=0.02$, obtained from Langevin Dynamics simulations.}
		\label{sisf_sim}
	\end{figure}

    This observation indicates that the increasing anisotropy of the confining potential drives the displacement distribution of the particles towards non-Gaussian behavior. At the same time the transfer of signal energy occurs from SVD mode 2 to mode 1. The signifcant transfer of signal energy from the circulation (mode 2) to the breathing-type pattern (mode 1) and the concomitant rise in non-Gaussianity is reminiscent of slowing down of particle dynamics in spatially constrained many body systems due to coupling between collective modes of the system \cite{lang2012mode}. We therefore, plot the SISF at four different values of $\alpha$ as shown in Fig. \ref{sisf_sim}. Here also we have performed average over a large number of random independent initial conditions to supress the statistical flucutations. At smaller confinement anisotropy (Fig. \ref{sisf_sim}(a), \ref{sisf_sim}(b) and \ref{sisf_sim}(c)), the SISF shows a decaying trend, but it does not decay to zero within the experimental time interval for which the function is plotted. However, for the larger confinment ansitropy (Fig. \ref{sisf_sim}(d)),  the structural relaxation is completely arrested similar to the experimental result (Fig. \ref{sisf_expt}(d)). 
    
    {It has been shown that, even for ergodic confined systems, finite-time trajectory averages need not coincide with ensemble averages, and substantial trajectory-to-trajectory fluctuations may persist \cite{jeon2012inequivalence}.} In the present study, at weaker confinement anisotropy (larger $\alpha$), the fluctuations among trajectories corresponding to different initial conditions are large. The simulation results are obtained by averaging over large number of random independent initial conditions, thereby sampling a much larger statistical ensemble. As a result, the initial condition averaged dynamical simulation observables, such as SISF and MSD can differ significantly from the experimentally measured single trajectory results. However, for stronger confinement (smaller $\alpha$) there occurs small fluctuations about mean dynamical observables. The dynamics then become less dependent on the particular initial condition, and fluctuations between different realizations are reduced. As a result, the experimental trajectory and the initial condition-averaged simulation exhibit more comparable dynamical behavior, leading to improved agreement between them. 
	\section{Summary and Conclusions}\label{summary}
	In summary, we investigated the effect of changing confinement anisotropy on the transport of particles in a finite cluster of charged dust particles in complex plasma. First the particle trajectories were decomposed into spatiotemporal modes using Singular Value Decomposition and it is shown that the decreasing dominance of mode 2 and an increasing dominance of mode 1 with increasing confinement anisotropy leads to an enhancement of non-Gaussianity of the particles' transport in the cluster. It was previously seen that the anisotropy parameter ($\alpha$) changes non-linearly with the change in electrode separation and is very sensitive to it \cite{sahu2025confinement}. Increasing confinement anisotropy induces coupling between the two dominant modes, resulting in an exchange of their character at the highest anisotropy. The transition of the transport of particles from a jump dominated transport at weak anisotropy to a more continuous and localized motion at stronger anisotropy happens with respect to the gradual change in the channel width, without any change in the background plasma parameters. The structural relaxation of the anisotropically confined cluster was also investigated via the calculation of autocorrelation of microscopic particle density. Both experiment and simulation results indicate that the strucutral relaxation is arrested at sufficiently stronger confinement anisotropy. However, at weaker anisotropy, the initial condition averaged simulation results exhibit larger deviation from experimental results obtained from single-trajectories. The discrepancy between the simulation and experimental results at weaker anisotropic confinement arises mainly due to the enhanced sensitivity of single experimental trajectories to initial conditions. At stronger confinement anisotropy, the fluctuation among different realizations decreases which leads to closer agreement of the simulation and experiemntal results.
    Our work also highlights the use of Singular Value Decomposition in characterizing particle dynamics in finite anisotropically trapped system of particles. 
	\section*{acknowledgement}
    A.S. acknowledges the Indian National Science Academy for the INSA Honorary Scientist position. All the computations were performed on the HPC cluster ANTYA  at the Institute for Plasma Research, Gandhinagar, India.
    \section*{Conflict of interest}
	The authors have no conflicts to disclose.
    \section*{Ethics Approval}
    No experiments on animal or human subjects were used for the preparation of the submitted manuscript.
	\section*{Data Availability}
    The data that support the findings of this study are available from the corresponding author upon reasonable request.
	
	\appendix
    \section{Singular Value Decomposition (SVD) Technique}\label{svd}
    The Singular Value Decomposition (SVD) is a powerful technique used to identify spatiotemporal patterns in scalar or vector fields. It decomposes the space-time signal into a set of orthogonal spatial and temporal modes. In experiments on many-body systems, various physical quantities-such as temperature, density, or electric and magnetic field —are typically measured in discrete spatial locations and time points. The SVD framework can be readily applied to such signals to identify the dominant coherent structures governing the dynamics.
    
    Unlike Fourier transform or  wavelet transform that rely on predefined basis functions, SVD is an entirely data driven technique. It doesn't assume any functional form of the basis functions a priori; instead, it determines the orthogonal basis from the dataset. Consequently, SVD is particularly well suited to study systems where the  inherent modes can evolve with external control parameters. SVD assigns weights to the orthogonal modes which are called the Singular Values, so that each mode has a certain dominance over the dynamics. Modes having larger singular values have a larger fraction of the signal energy and determine the dominating dynamical structures of the system.

    For a signal measured at multiple space and time points, the SVD can be mathematically expressed as,
    \begin{equation}
        Z(i,j)=\sum_{k=1}^{P} \sigma_k \phi_k(i) \psi_k(j),
    \end{equation}
    where $i$ and $j$ respectively denotes the space and time index. 
    They follow the orthogonality condition,
    \begin{equation}
        \sum_{i=1}^{N} \phi_p(i)\phi_q(i)=\sum_{j=1}^{M} \psi_p(j)\psi_q(j) = \delta_{pq}.
    \end{equation}
    In the above equation, $N$ and $M$ denote the total number of spatial and temporal sampling points in the dataset, respectively.
    The spatial eigenfunctions $\phi_k(i)$ and the temporal eigenfunctions $\psi_k(j)$  represent the orthogonal spatial and temporal modes and are commonly called the $topos$ and $chronos$ respectively. $\sigma_k$ denotes the singular value of the $k$th mode.
    In our experiment, the particle positions are recorded at definite intervals of time. So, we get the time series of $x-$ and $y-$ positions of the particles. Using the positions of the particles we first construct the data matrix $Z$ as follows :
    \begin{equation}
    Z=
    \begin{pmatrix}
    x_{1,1} & x_{1,2} & ... & ...&x_{1, M}\\
    x_{2,1} & x_{2,2} & ... & ...&x_{2, M}\\
    : & : & : & : & : \\
    : & : & : & : & : \\
    x_{N,1} & x_{N,2} & ... & ...&x_{N, M}\\
    y_{1,1} & y_{1,2} & ... & ...&y_{1, M}\\
    : & : & : & : & : \\
    : & : & : & : & : \\
    y_{N,1} & y_{N,2} & ... & ...&y_{N, M}\\
    
    \end{pmatrix}.
    \end{equation}

     We now perform SVD on the $Z$  matrix as follows,
    \begin{equation}
    Z=USV^{T}.
    \end{equation}
    The matrix $U$ contains the $topos$  and the matrix $V$ contains the $chronos$. The matrix $S$ contains the singular values corresponding to the different modes.
    The relative weight of the $i$th mode is calculated according to the following formula:
	\begin{equation}
		e_i = \frac{\sigma_i^2}{\displaystyle \sum_{j=1}^{2N} \sigma_j^2}.
	\end{equation}
    This quantity captures the fraction of signal energy contributed by the $i$th mode to the overall dynamics and has properties of a probability distribution \cite{dudok1994biorthogonal}.

	\bibliographystyle{aipnum4-2}
\bibliography{newpaper,mona_references}

\end{document}